\documentclass[aps,pra,pdf,superscriptaddress,amsmath,amssymb,amsfonts,twocolumn,showpacs,nofootinbib]{revtex4-1}

\usepackage{amssymb}
\usepackage{eqnarray,amsmath}
\newcommand{\abs}[1]{\left| #1 \right|} 
\usepackage{graphicx}
\usepackage{epsfig}
\usepackage{dcolumn}
\usepackage{bm}
\usepackage{braket}
\usepackage{amsmath}
\usepackage{physics}
\usepackage{csquotes}
\usepackage{mathtools}
\usepackage{graphicx,color,xcolor,colortbl}

\usepackage[colorlinks=true,
linkcolor=blue,
filecolor=blue,      
urlcolor=blue,
citecolor=blue]{hyperref}

\usepackage[normalem]{ulem}

\newcommand{\blue}[1]{{\color{blue}#1}}

\definecolor{Gray}{gray}{0.85}
\definecolor{LightCyan}{rgb}{0.88,1,1}

\newcolumntype{a}{>{\columncolor{Gray}}c}
\newcolumntype{b}{>{\columncolor{white}}c}

\begin{document}

\title{Selective Rotation and Attractive Persistent Currents in Anti-Dipolar Ring Supersolids}

\author{K. Mukherjee}
\author{T. Arnone Cardinale}
\author{S. M. Reimann}
\affiliation{Mathematical Physics and NanoLund, Lund University, Box 118, 22100 Lund, Sweden}

\begin{abstract}
A repulsively interacting Bose-Einstein condensate  on a ring is well known to show persistent currents. For attractive interactions, however, a bound state may form that renders the rotation classical. Here we show that in a multiply-connected confinement, the strong in-plane attraction of an {\it anti-dipolar }condensate can form stacks of ring-shaped droplets which may coherently overlap to form a supersolid 
along the azimuthal symmetry axis of the system. Intriguingly, the functional behavior of the energy-angular momentum dispersion of the anti-dipolar ring condensate differs from that of a usual repulsive superfluid. The periodic maxima between persistent flow and the non-rotating ground state flatten significantly and the typical pronounced cusps in the energy dispersion also occur in the rotationally symmetric supersolid state.  A weak link results in the reduction of this minimum, shifting it to smaller angular momenta. With an asymmetric link potential one can selectively induce superfluid and rigid-body rotation in different layers within the same system. This intriguing setup offers new perspectives for atomtronics applications.
\end{abstract}
\date{\today}

\maketitle
\textit{\blue{Introduction.}}
A macroscopic coherent state with off-diagonal order~\cite{Gross1957, *Yang1962, *Leggett1999_SF} in a multiply-connected 
geometry may carry a persistent current (PC) with quantized angular momentum -- one of the hallmarks of superfluidity. 
PCs occur in superconductors~\cite{Deaver1961}, metallic rings~\cite{Levy1990, *Mohanty1999}, exciton-polariton systems~\cite{Sanvitto2010, *Lukoshkin2018} and liquid helium~\cite{Avenel1985},  and 
toroidally trapped Bose-Einstein condensates (BECs)~\cite{Ryu2007,Ramanathan2011_prl,Moulder2012,Wright2013_prl,Beattie2013,Murray2013,Eckel2014,Jendrzejewski2014_prl,YGuo2020,deGoerdeHerve2021}. 
Recent works also have provided evidence for supercurrents  with paired fermionic $^6$Li~\cite{cai_2022_fermions,*Delpace_2022_fermions}. 
Ultra-cold atomic quantum gases in general are well-known for exceptional experimental tunability~\cite{kohler2006, *Bloch2008, *chin2010}, enabling new concepts for quantum sensing and \enquote{atomtronic}  applications~\cite{Amico_2017, *amico_roadmap_2021, *Pepino2021_atomtronics2021, *Luigi_coloquim_2022}. 

Stirring a toroidal BEC with a laser beam 
introduces a potential barrier that locally suppresses the bosonic density forming a so-called \enquote{weak link} (WL)~\cite{Ramanathan2011_prl, Wright2013_prl, Eckel2014, Jendrzejewski2014_prl}. It allows a controlled mixing of different angular momentum states and is used as a basic operating scheme in atomtronic quantum interference devices (AQUIDs)~\cite{Matveev2002, solenov2010, Minguzzi_2014_prl, Mathey_2016, amico_2018_readout, Kiehn_2022, Polo_2022_attraction}. 
PCs have been studied in single~\cite{Svistunov2000, *Tempere2001,*Karkkainen2007, 
*Mateo2015,*Yakimenko2015,*Polo2019,*Bland2020doublering,*Bland2022persistent, *Giacomo_2023} and multicomponent~\cite{Smyrnakis_mixture_2009, *Malet_2010, *Zaremba_mixtur1_2013, *Zaremba_mixture2_2013, *Yakimenko_spinor_2013,*Abad2014,*Smyrnakis2014,*Wu_mixture_2015, *White_spin-orbit_2017} contact-interacting as well as 
dipolar BECs~\cite{Abad2010,Abad2011,Malet_2011, Karabulut2013,Tengstrand2021,Tengstrand2023}. 
A BEC with short-range attractive interactions, however,  is stable only below a certain atom number~\cite{hullet2000,kanamoto_2003_stable, Kavoulakis2003}. Only few studies so far examined the rotational properties of attractive BECs in the regimes of uniform~\cite{ueda1999, kanamoto_attarctive_rotation1_2003} or bright quantum-solitonic densities~\cite{Naldesi_2022_scipost, Polo_2022_attraction, chergui_2023}.
\\
\begin{figure}
	\centering
	\includegraphics[width = 0.45\textwidth]{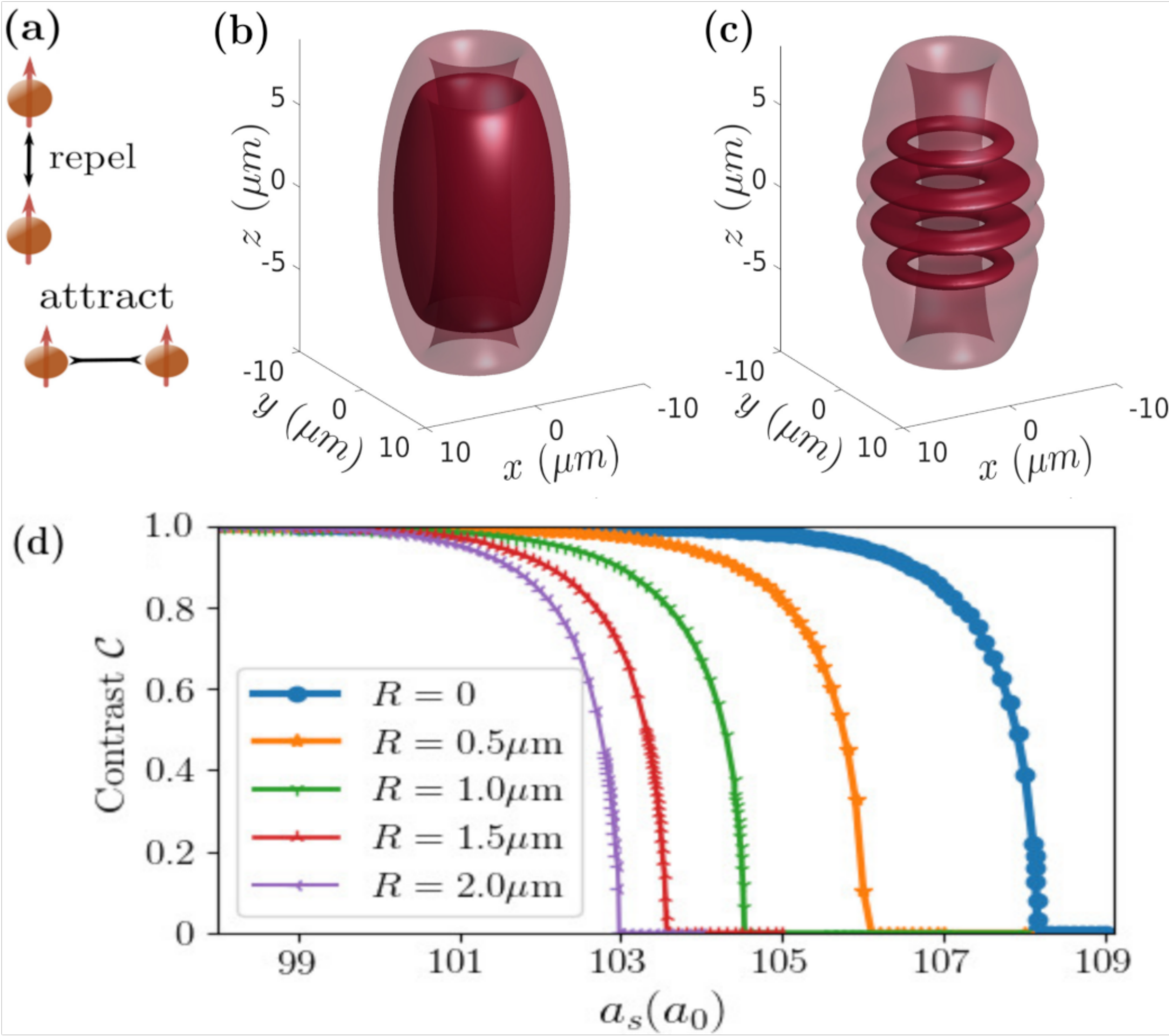}
	\caption{(a) Schematic illustration of interactions between anti-dipoles.  (b) Isosurfaces of the three-dimensional (3D) density, depicting  a non-modulated  superfluid  and (c) supersolid state, forming in a 3D elongated confinement with a central hole of size $R=1.5 \rm \mu m$, and scattering length $a_s = 105a_{0}$, and $a_s=103a_{0}$, respectively. The isosurfaces are taken at $50\%$ and $5\%$ of the maximum density. (d) The density contrast $\mathcal{C}$ as a function of  $a_s$ for different size $R$. The system is composed of $N = 5 \times 10^{5}$ atoms with dipolar length $a_{dd} = -65.5a_{0}$.} \label{figure1}
\end{figure}
Here, we uncover a new and rich scenario for PCs in the attractive regime, leveraging the specific features of {\it anti-dipolar} 
BECs~\cite{Tang2021} where the constituting magnetic dipoles attract in a side-by-side configuration, but repel in the head-to-tail 
arrangement~\cite{Giovanazzi2002, *Prasad2019, *Baillie2020} (see Fig.~\ref{figure1}(a)). 
This, in fact, is just {\it opposite} to the interaction in a regular dBEC~\cite{Baranov2008, Lahaye2009, Baranov2012}. 
Nevertheless, the system can exhibit both diagonal and off-diagonal order simultaneously, resulting in an intriguing supersolid state (SS) that 
manifests here as a stack of flat condensate slices piling up along the weakly confined direction, say, the 
$z$-direction~\cite{WenzelPhD,Mukherjee_stack_2023}. In a multiply-connected geometry, the supersolid resembles a stack of coupled rings, with a homogeneous density extending across the entire azimuth and phase-coherently overlapping along $z$,  as shown by the density isosurfaces in Fig.~\ref{figure1} (c). 
Supersolids~\cite{Yang1962, *Andreev1969, *Chester1970, *Leggett1970,*Pomeau1994} were long sought in $^4$He~\cite{Kim2004a,*Balibar2010, *Kim2012}, but evidence came from experiments with ultra-cold atoms~\cite{Leonard2017a, *Leonard2017b, *Li2017, Ostermann2016}, in particular from dBECs of dysprosium~\cite{Tanzi2019,Boettcher2019} and erbium atoms~\cite{Chomaz2019}. These supersolids are formed by coherently overlapping arrays of droplets~\cite{ saito2016,Kadau2016, Ferrier2016, Chomaz2016, Chomaz2018} stabilized by energy corrections beyond mean-field~\cite{Lima2011, Lima2012, Petrov2015, Petrov2016, Wachtler2016a, Bisset2016} (see also the review~\cite{Chomaz2022} and references therein). Anti-dBECs~\cite{Tang2021} in this context however still remain a largely unexplored territory~\cite{WenzelPhD,Mukherjee_stack_2023, Arazo2023, Kirkby2023}. 

In this Letter, starting from anti-dipolar toroidal supersolids -- which are interesting in their own right -- we demonstrate 
how new and  intriguing properties of PCs arise from the in-plane attractive and inter-plane repulsive interactions,
substantiated by the energy-angular momentum dispersion relation and controllable through WL symmetry.   
Strong anti-dipolar interactions give rise to a dispersion relation that is significantly different from the usual functional form for a repulsive superfluid. The rotational properties can be engineered by asymmetric WLs, making it possible to induce both superfluid and rigid body rotation within the same system, prominently showcased in the dynamical properties of the system.

\textit{\blue{Model and method.}} In an anti-dipolar setting, the long-range interaction differs from the usual dipolar case just by a minus sign,  reversing its role. Just like dBECs, also anti-dBECs are well approximated by the non-local extended Gross-Pitaevskii 
equation~\cite{pethick, string, Chomaz2022},
 \begin{eqnarray}\label{eGPE}   
 & i\hbar \frac{\partial \psi(\textbf{r},t)}{\partial t}  =  \bigg[-\frac{\hbar^2}{2M}\nabla^2 + V(\textbf{r}) + g \abs{\psi(\textbf{r},t)}^2- \frac{3}{4 \pi} g_{\rm dd}\times \nonumber\\&  \int d\textbf{r}^{\prime} \frac{1 -3\cos^{2}\Theta}{\abs{\textbf{r} - \textbf{r}'}^3}\abs{\psi(\textbf{r}^{\prime},t)}^2 + \gamma(\epsilon_{dd})\abs{\psi(\textbf{r},t)}^3 \bigg] \psi(\vb{r},t). 
 \end{eqnarray}
The external toroidal potential with an additional WL is expressed as  $V(\vb{r}) = M\omega^2_{0}[r^2 + \lambda^2 z^2]/2 + V_l  e^{-\theta^2/w^2_l} + V_{0}e^{-r^2/(2R^2)}$, where $M$ is the atom mass, $\lambda = \omega_z/\omega_0$  is the ratio of the trapping frequency in the $z$-direction to that in the radial plane, and the Gaussian creates the central hole with radius $R$. The parameters $V_l$ and $w_l$ control strength and thickness of the WL  constricting the ring confinement around $\theta =0$. 
The short-range repulsive contact interaction,  $g = 4 \pi \hbar^2a_s/M$, is determined by the scattering length $a_s$. 
Moreover, $g_{\rm dd} = 4\pi\hbar^2a_{\rm dd}/M$, where $a_{\rm dd}= \mu_{0}\mu^2_{m}M/24\pi \hbar^2$ represents the dipolar length corresponding to the maximum anti-dipolar interaction.
The last term in Eq.~\eqref{eGPE} is the so-called Lee-Huang-Yang (LHY) correction~\cite{Fisher_2006, Lima2011, Lima2012}
 where $\gamma(\epsilon_{\rm dd}) = \frac{32}{3}g \sqrt{\frac{a^3_s}{\pi}} \left(1+\frac{3}{2}\epsilon_{\rm dd}^2\right)$. The dimensionless parameter $\epsilon_{\rm dd} = a_{dd}/a_s$ quantifies the relative strength of the DDI as compared to the contact interaction. For a rotating WL, we consider a reference frame that moves with it, which adds a term $-\Omega L_{z}\psi$ to the right-hand side of Eq.\eqref{eGPE}, where $\Omega$ represents the rotation frequency and $L_{z} = {\rm i} \hbar (y\partial_x - x \partial_y)$~\cite{Fetter_2009}. 
 The solution of Eq.~\eqref{eGPE} is
obtained by employing the the split-step Fourier method in imaginary time to determine the ground
states at fixed angular momentum $L=L_0$ as in ~\cite{Komineas_2005, karkkiainen_persistent_2007},  
and in real time to monitor the dynamical properties.  We remark that obtaining the stationary state solution of Eq.~\eqref{eGPE} can be challenging due to the presence of numerous closely spaced local minima in the energy surface. This necessitates thorough sampling across various initial conditions to identify the probable lowest-energy solutions.

In the following, we consider an anti-dipolar dBEC of $ ^{164}$Dy with $N = 5 \times 10^{5}$
atoms. The frequencies of harmonic potential are set to $\omega_0/(2\pi) = 100 \rm Hz$ and $\omega_z/(2 \pi) = 50 \rm Hz$, resulting in an elongated geometry along the z-axis.

\textit{\blue{Different phases of the toroidal anti-dBEC.}} First, we determine the different ground state phases by evaluating density contrast, $\mathcal{C} = ( n_{\rm max} - n_{min} )/(n_{max} + n_{min})$,  as a function of the scattering length $a_s$. Here $n_{\rm max}$ and $n_{\rm min}$ are maximum and minimum densities, respectively. 
$\mathcal{C} = 0$ relates to a non-modulated density profile, as shown in Fig.~\ref{figure1}(a),  essentially representing the shape of the underlying confinement potential. The range $0 < \mathcal{C} \le 1$ signifies an emergence of spontaneous density modulation along the $z$-direction. This affirms the formation of stacks of droplet rings, which may either be linked by a background superfluid ($0 < \mathcal{C} < 1$), forming a toroidal SS  [Fig.~\ref{figure1}(b)], or be isolated from each other ($\mathcal{C}=1$) when the superfluid background vanishes. In Fig.~\ref{figure1}, we illustrate how $\mathcal{C}$ varies with $a_s$ for different sizes of the central hole of radius $R$.
A reduction in $a_s$ amplifies in-plane attraction and axial repulsion, yielding $\mathcal{C} > 0$. 
It is intriguing how the anisotropic long-range interactions uniquely facilitate the formation of ring condensates with an effective in-plane attraction. In contrast, regular  BECs with attractive contact interactions are prone to form a density blob, or undergo a collapse~\cite{hullet2000}. \par 

The in-plane long-range attraction for the anti-dBEC is most pronounced at the plane center close to the $z$-axis, and is thus effectively diminished by introducing a central potential barrier pushing the atoms to larger radii. Lowering $a_s$ (compared to $R=0$ case) becomes necessary to render the system in a density-modulated state. This is evident from Fig.~\ref{figure1}(a), where the transition point for the modulating state shifts towards lower $a_s$ with increasing $R$. Additionally, a sufficient amount of short-range repulsive interaction is needed to submerge the stacks of droplet rings within a dilute superfluid and thereby produce the SS state. 
Attaining the SS state over a broad range of $a_s$ may necessitate increasing the particle number in the system, as reflected by the relatively large value chosen here. Nevertheless, the results shown below hold for the SS ring phase independent of the specific $N$. 
\begin{figure}
	\centering
	\includegraphics[width = 0.49\textwidth]{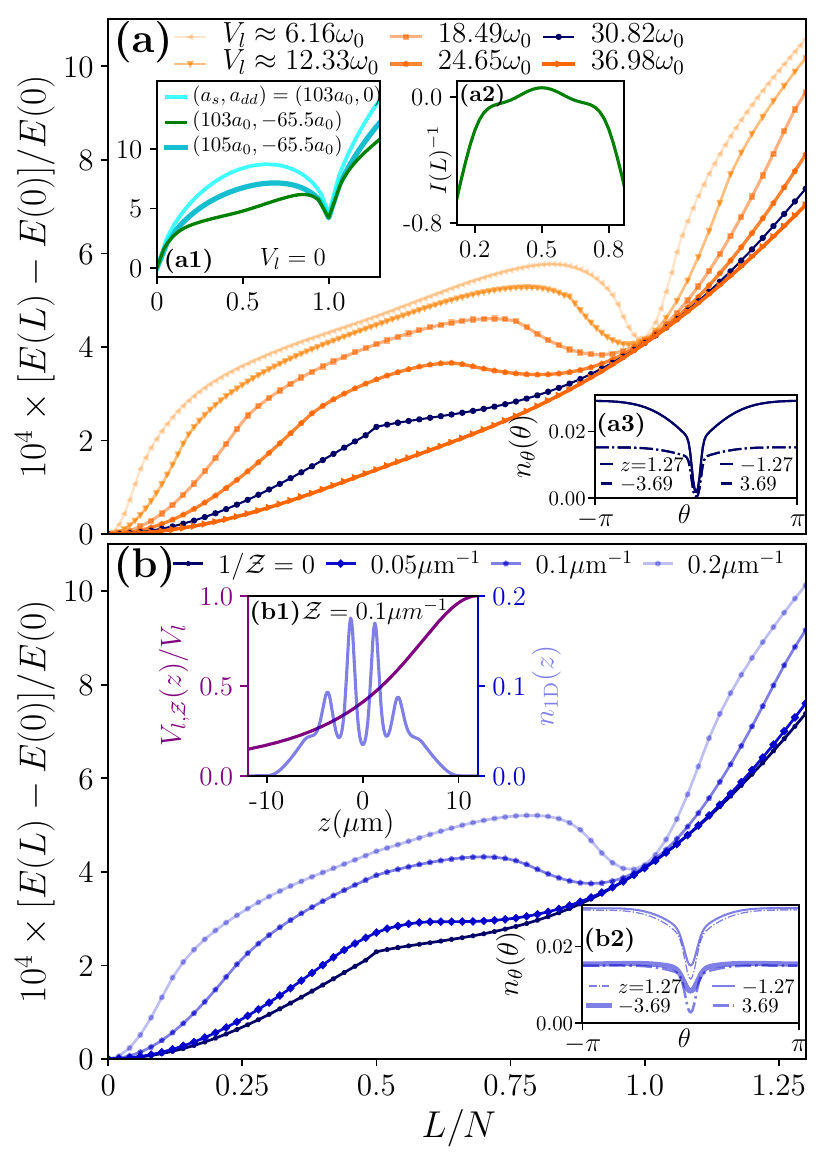}
	\caption{(a) Ground state energy dispersion $E(\ell )$ relative to the non-rotating energy $E(0)$ of an anti-dipolar ring SS, as in Fig.~\ref{figure1}(c), as a function of angular momentum $L$ for different values $V_{l}$ (see the legends) with a WL at $\theta = 0$ that is symmetric with respect to $z$. Inset (a1) as in (a), but for different scattering length $a_s$ at $V_{l}=0$. Inset (a2) shows the inverse moment of inertia $I^{-1}(\ell )$ for $a_s = 103a_{0}$. (b) As in (a) but for an asymmetric WL  characterized by the parameter $\mathcal{Z}$ (see legend). The variation of the WL potential $V_{l,\mathcal{Z}}$ across $z$ along with the integrated one-dimensional density $n_{\rm 1D}(z)$ is shown in Fig.~(b1) for $\mathcal{Z}=0.1\mu m^{-1}$ and $V_{l}= 30.82 \omega_0$. (b2) The integrated one-dimensional density $n_{\theta}$ as function of azimuthal angle $\theta$ calculated at different $z$ corresponding to the location the of the density maxima as in (b1).  } \label{figure2} 
\end{figure}

\textit{\blue {Rotation with a symmetric weak link.}} Now we take a look at the rotational properties  of the anti-dipolar toroidal supersolid, for various $V_l$ and a fixed $w_l=0.1$. Representative azimuthal variations of density,  $n_{\theta}(\theta) = \int d\rho \abs{\psi(\rho, \theta, z_1)}^2$, with $\rho = \{x, y\}$, 
at the position of droplet rings, $z_1$, are shown in Fig.~\ref{figure2}(a3) for $V_l =30.82 \omega_0$. 
Here, the WLs are symmetric with respect to z. In a usual dipolar supersolid, the  density has a tendency to increase adjacent to the WL due to the repulsive interaction between parallel dipoles~\cite{Tengstrand2023}. 
Consistently, in the anti-dipolar case it is maximal at $\theta = \pm \pi$. 
In Fig.~\ref{figure2}(a), we show the ground state energy, $E(\ell )$, where $\ell =L/N$, for different $V_l$.  The system always exhibits a global minimum at $L=0$. 
It is intriguing to explore if there are additional local minima in the dispersion indicative of metastable states that could support PCs. When $V_l$ is sufficiently strong (e.g., $V_l = 36.97 \omega_0$), the density at the position of the WL is fully depleted. The dispersion relation in this case forms a smooth function of $\ell $ with non-negative curvature. Consequently, the system cannot carry any PC and instead only rotate like a rigid body with the WL. Reducing $V_l$ to $30.82\omega_0$, some density accumulates at the WL position, and the dispersion relation adopts the well-known form of intersecting parabolae,  however still lacking a local minimum. With a continued decrease in $V_{l}$, the curvature of the second parabola becomes more pronounced, and a second minimum emerges at $\ell < 1$, enabling the system to sustain a PC. \par

Notably, the strong in-plane attractive interaction of the anti-dBEC results in a reduction of the curvature in the dispersion curve between the minima. This is particularly prominent as $V_l$ decreases, causing a substantial flattening of the dispersion curve. 

To assess the impact of attractive interaction, we show in Fig.~\ref{figure2}(a1) the dispersion without WLs~(i.e., $V_l = 0$)  for different interaction parameters, comparing it also to the usual superfluid case with isotropic contact interactions only. 
For the anti-dipolar superfluid regime, similar to the usual non-dipolar superfluids, the dispersion has the familiar shape with a concave downward cusp at $\ell $=1.  It fundamentally changes its functional form for stronger attraction when $(a_s, a_{dd}) = (103a_{0}, -65.5a_{0})$; notice, in particular, the flattening in the range $0.25 < \ell < 0.75$.  This can be traced back to the moment of inertia~\cite{Leggett_2001_review} calculated as $I^{-1} = \partial^2 E/\partial \ell^2$  shown in Fig.~\ref{figure2}(a2). 
While the curvature of the dispersion typically is a positive constant for rigid-body rotation and is negative in the superfluid limit, 
here it mostly exhibits linear variation, being close to zero on the negative side, as well as positive values within $0.38 < \ell < 0.62$.
\begin{figure*}
	\centering
	\includegraphics[width = 0.999\textwidth]{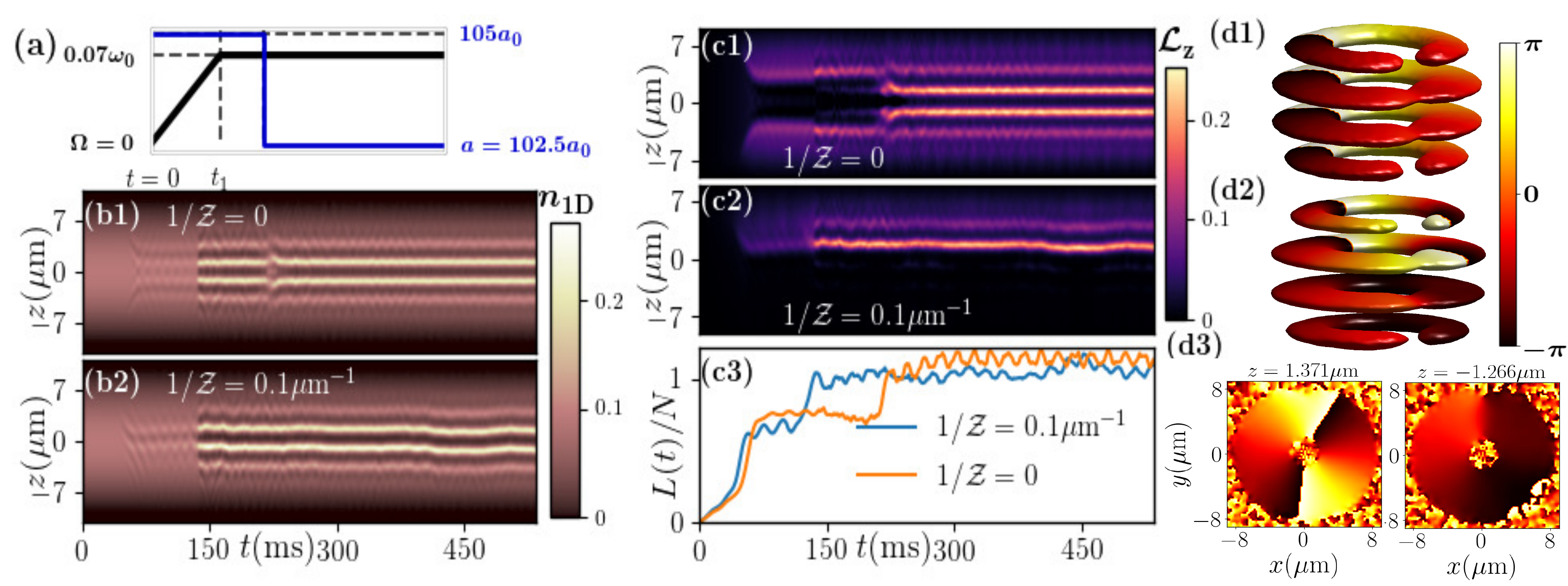}
	\caption{(a) Schematic of the dynamical protocol. We begin with an un-modulated non-rotating anti-dBEC at $a_s = 105a_{0}$, and gradually change the rotation frequency to its final value at time $t_1$. Then we suddenly decrease the scattering length corresponding to a modulated state at $t_2$. Panels (b2)-(b3)  show  the time evolution of the integrated density $n_{\rm 1D}(z)$, and (c1)-(c2) the integrated angular momentum density $\mathcal{L}_{z}(z) = (1/N)\int dx dy~\psi^{*}L_z\psi$ for a symmetric [(b1), (c1)] and asymmetric WLs [(b2), (c2)]. The corresponding time evolution of the total angular momentum is shown in (c3). The panels to the right show the three-dimensional density isosurface ($30\%$ of the maximum density) at $t=530 \rm ms$ for the (d1) symmetric and (d2) asymmetric WLs, with color indicating the phase. While the superfluid rotation is created through the entire system for the symmetric WL, the same occurs only for $z>0$ for the asymmetric WL. (d3) The corresponding phase profiles at $z=1.371 \mu m$, and $z = -1.266 \mu m$ showcasing superfluid and rigid-body rotation, respectively.} \label{figure3} 
\end{figure*}

\textit{\blue{Asymmetric weak link.}} Let us now investigate ways to maneuver the rotational response of the system with WLs that are not symmetric with respect to $z$. We here employ a WL potential with varying amplitude and thickness around $\theta = 0$ by performing the  transformation $V_{\rm l }(z) \rightarrow V_{l}/\sqrt{f(z)} \equiv V_{l, \mathcal{Z}}(z)$,  and $w_{l}(z) \rightarrow w_{l}\sqrt{f(z)}$ where $f(z) = \sqrt{1 + (z-z_{R})^2/\mathcal{Z}^2}$, and the parameter $\mathcal{Z}$ quantifies the degree of asymmetry.  The functional form of $f(z)$ is motivated by the familiar mathematical expression determining the radius of a Gaussian laser beam with waist located at $z_R$, which would also well describe an experimental setup for the stirring. 
The axial variation of the $V_{\l, \mathcal{Z}}$(z) and integrated one-dimensional density $n_{\rm 1D}(z) = \int dx dy \abs{\psi}^2$ are shown in Fig.~\ref{figure2}(b1) for $\mathcal{Z}=0.1 \rm \mu m^{-1}$ and $z_R = 12 \rm \mu m$. It becomes evident from the $n_{\rm 1D}$(z) that the particles are propelled towards lower $V_{l, \mathcal{Z}}(z)$, increasing the density dip at $\theta = 0$ for $z>0$  compared to $z<0$; see also the azimuthal variation of the density slices, $n_{\theta}$ in Fig.~\ref{figure2}(b2), calculated at the position of droplet rings $z=z_1$ corresponding to the maxima of $n_{1D}(z)$ in Fig.~\ref{figure2}(b1).
Focusing on the parameters $a_s=103a_{0}$ and $V_{l} = 30.82\omega_0$, in Fig.~\ref{figure2}(b), we show the dispersion relation for different $\mathcal{Z}$. Noticeably, the appearance of the second minimum and thus the existence of a PC is favoured when introducing the asymmetry. In particular, for larger asymmetry, the dispersion contains both a wide flattened region, stemming from more azimuthaly symmetric droplet rings located next $z=0$, and a wide opening of the underlying parabola at the second minimum, originating from the the large value of $V_{\rm l, \mathcal{Z}}$ at $z=3.69 \rm \mu m$.  Consequently, as we further outline below, this gives rise to the intriguing possibility for a vortex to penetrate droplet rings located at $z > 0$, but {\it not} through the ones located at $z < 0$, making a spatially selective PC possible. 

\textit{\blue{Time-dependent simulation.}} 
To further elucidate the rotational properties of droplet rings, we now study their dynamics. We start with a non-modulated ground state at $a_s = 105a_{0}$ [Fig.~\ref{figure1}(b)]. Subsequently, we linearly increase the rotation frequency $\Omega$ to its final value of $0.07 \omega_0$ over a duration of $t_1 = 59 \text{ms}$. Following this, we abruptly decrease the scattering length to its final value, $a_s=102.5a_0$, at $t_2 = 135 \text{ms}$ [see the protocol in Fig.~\ref{figure3}(a)] to induce a modulated density state, consisting of droplet rings.
The time evolution of $n_{\rm 1D}(z,t)$ is shown in Fig.~\ref{figure3}(b1) and Fig.~\ref{figure2}(b2) for $\mathcal{Z}=0$ and $\mathcal{Z}=0.1 \mu m^{-1}$, respectively. The densities of the middle two droplets are more pronounced for $\mathcal{Z}=0.1 \mu m^{-1}$, with the superfluid density at $z=0$ being almost non-existent. As illustrated in Fig.~\ref{figure3}(c3), the time evolution of the angular momentum per particle indicates that the PC of value $\ell  \approx 1$ can indeed be maintained by the dynamically generated droplet rings. 

An intriguing question that arises here is, if and how the individual droplet rings make a contribution to the PC.
We thus calculate the angular momentum density, $\mathcal{L}_{z}(z) = (1/N)\int dx dy~\psi^{*}L_z\psi$, along $z$.  Remarkably, while the superfluid rotation is collectively shared by all droplets when $\mathcal{Z}=0$, the same occurs only for the ones located at $z > 0$, for $\mathcal{Z}=0.1 \mu m^{-1}$. The reason for this behavior is as follows: While a vortex line permeates through all droplet rings for $\mathcal{Z}=0$, it can only penetrate those at $z > 0$ due to a smaller WL density stemming from axially asymmetric potential bumps. Here, the vortex line exits the condensate via the null density region at $z=0$. This observation is further supported by examining the phase of the droplet rings.  The density isosurafaces ($30 \%$ of the maximum density) at $t=534 \rm ms$, for $\mathcal{Z}=0$ and $\mathcal{Z}=0.1 \rm \mu m^{-1}$, are displayed in Fig.~\ref{figure3}(d1) and  Fig.~\ref{figure3}(d2), respectively. The colors indicate the corresponding values of phase. The typical phase jump of $2\pi$, corresponding to unit angular momentum, is clearly visible across all droplet rings for $\mathcal{Z}=0$. In contrast, for $\mathcal{Z}=0.1 \mu m^{-1}$, a double winding of the phase around the central region can be observed for the top two droplet rings. However, the phase is more uniform for the bottom two rings (see also the right panel of Fig.~\ref{figure3}(d3)), indicating that the vortex line does not pass through them, resulting in an absence of PC. Notably, since the total angular momentum per particle of the entire system is roughly one unit, mainly contributed by the particles in the upper half-portion of the condensate, each particle there carries two units of angular momentum. This is evident from the phase profile calculated at $z = 1.371 \rm \mu m$ [Fig.~\ref{figure3}(d3)]. In this manner, a unique system of stacked droplet rings is created, where a few rings maintain a PC and act like a superfluid, while others behave like a rigid body under rotation.\par
\blue{\textit{Conclusions}.}
In conclusion, we have shown how a supersolid anti-dBEC enables selective PCs in the presence of strong interatomic attraction. While the observation of PC in short-range attractive BECs remains inconclusive, the anisotropic interaction within an anti-dBEC serves to prevent collapse of the attractive condensate, creating favorable conditions also for strong attractions. The formation of a supersolid stack of droplet rings opens up further possibilities  to maneuver the rotational properties by introducing an asymmetric WL. This allows for superfluid rotation in certain droplet rings that permit the passage of vortex lines, while others, which resist such passage, exhibit rigid body rotation. It will be intriguing to further explore how to control the selective PC  
for different numbers of ring droplets in the supersolid.  
Exploiting such controllability of this novel rotational behavior will allow  to construct new types of AQUIDs and parallel circuits with complementary properties~\cite{amico_roadmap_2021}. In this context, it will be highly relevant to examine the collective excitation spectra~\cite{Hertkorn2019, Hertkorn2021densityfluctuations}, finite temperature impact~\cite{Sanchez2022}, Josephson oscillations~\cite{Albiez2005_josephson, mistakidis_dy_2024} and anti-dipolar mixtures for this intriguing new type of supersolid. 
\par 
\textit{Acknowledgements.} This work was financially supported by the Knut and Alice Wallenberg Foundation and the Swedish Research Council. Discussions with P. St\"{u}rmer, L. Chergui, and M. Schubert are gratefully acknowledged. 
\bibliographystyle{apsrev4-1.bst}
\bibliography{reference.bib}
\end{document}